\documentclass[pof,superscriptaddress,singlecolumn]{revtex4-1}
\usepackage[dvips]{graphicx}
\usepackage{amssymb,amsfonts,amsmath,color,psfrag}

\newcommand{\bs}{{\bf {s}}}
\newcommand{\br}{{\bf {r}}}

\newcommand{\bv}{{\bf {v}}}

\newcommand{\etal}{{\it et al.}~}

\newcommand{\bnabla}{\boldsymbol{\nabla}}

\begin{document}

\title{Vortex line density in counterflowing He II with laminar and turbulent normal fluid velocity profiles}

\author{A. W. Baggaley}\email[]{andrew.baggaley@gla.ac.uk}
\affiliation{
School of Mathematics and Statistics, University of Glasgow,
Glasgow, G12 8QW, UK
}

\author{S. Laizet} \email[]{s.laizet@imperial.ac.uk}
\affiliation{Turbulence, Mixing and Flow Control Group, Department of
  Aeronautics, Imperial College London, London SW7 2PG, UK}

\begin{abstract}
Superfluid helium is an intimate mixture of a viscous normal fluid,
with continuous vorticity, and an inviscid superfluid, where vorticity
is constrained to thin, stable topological defects.  One mechanism to
generate turbulence in this system is through the application of a
heat flux, so called thermal counterflow.  Of particular interest is
how turbulence in the superfluid responds to both a laminar and
turbulent normal fluid in the presence of walls.  We model
superfluid vortex lines as reconnecting space curves with fixed
circulation, and consider both laminar (Poiseuille) and turbulent
normal fluid flows in a channel configuration.  Using high
resolution numerical simulations we show that turbulence in the normal
fluid sustains a notably higher vortex line density than a laminar
flow with the same mean flow rate. We exam Vinen's relation, $\sqrt{L}=\gamma v_{ns}$, between
the steady state vortex line density $L$ and the counterflow velocity $v_{ns}$.
Our results support the hypothesis that transition to turbulence in the normal fluid is responsible for the TI to TII transition. 
We also consider the spectral
properties of fluctuations of the superfluid vortices, which show a
good agreement with previous experimental results.
\end{abstract}

\pacs{\\
Vortices in superfluid helium-4, 67.25.dk\\
Hydrodynamic aspects of superfluidity, 47.37.+q\\
Turbulent flows: channel flow, 47.27.nd
}

\maketitle

\section{Introduction}
\label{section:1}

Turbulence, the chaotic, disorder motion of fluids, poses tremendous
mathematical, physical and engineering challenges.  In recent years a
large body of work \cite{Skrbek12,VinenNiemela} has highlighted
similarities between the turbulence of superfluid helium~II (quantum
turbulence) and the turbulence of ordinary (viscous) fluids.  For
example, both experimental \cite{Maurer1998,Salort2010} and numerical
\cite{Nore1997,Araki2002,BaggaleyEPL} studies have shown that the
distribution of the superfluid kinetic energy is consistent with the
$k^{-5/3}$ Kolmogorov scaling observed in classical turbulence.  Other
experimental studies have also observed Kolmogorov's four-fiths law
\cite{SalortEPL}, and quasi-classical behaviour in the decay of
quantum turbulence \cite{SkrbekPRE2003,walmsley2008quantum}

The similarity is particularly striking as superfluid helium is unlike
an ordinary fluid.  It possesses a two-fluid nature, indeed the system
is an intimate mix of a viscous normal fluid component and an inviscid
superfluid component, which are coupled through mutual friction
\cite{Donnellybook}.  Perhaps of more interest is that due to the
constraints of quantum mechanics, the superfluid vorticity field is
not continuous.  Instead it is restricted to discrete vortex filaments
around which the circulation is fixed to the ratio of Planck's
constant and the mass of one helium atom.  The commonly held
definition of quantum turbulence is the motion induced by a tangle of these
quantised vortices.  Hence, quantum turbulence may provide a
simplified model, a `skeleton', where one can hope to learn more about
the complex mechanisms present in classical turbulence.

Whilst the similarities between quantum and classical turbulence are
striking, there are other forms of quantum turbulence which are unique
to this two-fluid system. One of these, thermal counterflow
\cite{Tough1982}, is the subject of this study.  This form of
turbulence is typically created by heating one end of a channel
containing superfluid helium.  The heat flux is carried from the
heater via the normal fluid alone and, due to conservation of mass,
the normal fluid and the superfluid move in opposite directions;
their velocity difference is proportional to the applied heat
flux. This was the setting for the earliest studies of quantum
turbulence \cite{Vinen1,Vinen2,Vinen3,Vinen4}, and still poses a
number of open questions, which continue to attract experimental
\cite{SkrbekPRE2003,BarenghiPRE2006} and theoretical studies
\cite{Adachi2010,Galantucci2011}.

It is important to stress that although counterflow has no classical
analogy, it does not necessarily mean that the study of counterflow
cannot offer deeper insights into phenomena in classical
turbulence. For example Skrbek \etal \cite{SkrbekPRE2003} emphasised
that the efficiency of turbulent heat transport in counterflow is
similar to that of turbulent thermal convection in a classical fluid.
Of particular interest in recent years has been the question of the
transition from laminar to turbulent flow in the normal fluid
component \cite{Guo2010}. One would expect that a deeper
understanding of this transition would be important when considering
the transition to turbulence in a classical fluid.

This transition to turbulence is also believed to be important in understanding the different
turbulent states observed in thermal counterflow.
In channels with a small (cross-sectional)
aspect ratio, Tough \cite{Tough1982}
categorised two turbulent regimes TI and TII, 
which exhibit strong difference in the scaling of the steady state vortex line density to the applied heat flux.
Stability analysis by Melotte\& Barenghi  \cite{Melotte1998} indicated that transition to turbulence in the normal fluid
was the underlying reason for the transition from the TI to TII regime.
Fully probing this transition is beyond the scope of this current study; our
goal here is to better understand differences in the structure of the
quantised vortices when the normal fluid is laminar (Poiseuille) and
turbulent. To this end we perform high resolution numerical
simulations using the vortex filament method and compare and contrast
statistical information about the steady state tangles at two
different temperatures.

In the next section we discuss our approach to numerically modelling
the evolution of quantised vortices
(section~\ref{section:2}). Section~\ref{section:3} is devoted to
details about the numerical modelling of a turbulent normal fluid.
Section~\ref{sec:results} summarises our findings, focusing in
particular on the structure of the vortex tangle, as well as
fluctuations of the vortex line density. Finally we close with brief
comments and conclusions in section~\ref{sec:conc}.

\section{The Vortex Filament Method}
\label{section:2}

Following Schwarz \cite{Schwarz1985}, we model quantum vortex
filaments as space curves $\bs(\xi,t)$ which move according to

\begin{equation}
\frac{d{\bf s}}{dt}=\bv_s+\alpha \bs' \otimes (\bv_n-\bv_s)
-\alpha' \bs' \otimes (\bs' \otimes (\bv_n-\bv_s)),
\label{eq:Schwarz}
\end{equation}

\noindent
where $t$ is time, $\alpha$ and $\alpha'$ are temperature dependent
friction coefficients \cite{Donnelly1998}, $\bs'=d\bs/d\xi$ is the
unit tangent vector at the point $\bs$, $\xi$ is arc length, and
$\bv_n$ is the externally applied normal fluid velocity.  We perform
all calculations in the frame of reference of the imposed superflow,
hence the total superfluid velocity $\bv_s$ reduces to the
self-induced velocity of the vortex filament, which is given by the
Biot-Savart law
\begin{equation}
\bv_s(\bs)=
-\frac{\Gamma}{4 \pi} \oint_{\cal L} \frac{(\bs-\br) }
{\vert \bs - \br \vert^3}
\times {\bf d}\br,
\label{eq:BS}
\end{equation}

\noindent
where $\Gamma=9.97 \times 10^{-4}~\rm cm^2/s$ is the quantum of
circulation and the line integral extends over the entire vortex
configuration $\cal L$.

In numerical studies of thermal counterflow
\cite{Schwarz1988,Adachi2010,Sherwin2012}, it is common to simplifying
the problem by taking a uniform velocity profile for $\bv_n$,
neglecting the important effect of boundaries. In addition no
numerical study has investigated the role turbulence in the normal
fluid may have on the system. In this study we take a simple analytic
Poiseuille profile, suitable for a laminar normal fluid in a
channel. This is contrasted with a turbulent velocity field,
obtained by numerically solving the incompressible
Navier-Stokes equations. Both velocity fields will be described in
more detail below, however it is important to state here that in all
cases the normal fluid is simply prescribed, ignoring any modification
of the flow due to the presence of the quantised vortices.  While this
is clearly sub-optimal, a dynamically self-consistent model would be
very complex and computationally expensive. Indeed an imposed normal
fluid is still the most common approach in the literature
\cite{Adachi2010,Morris2008,Kivotides2001a,Kivotides2006}, and a
better understanding of key physical processes can be garnered from
such an approach.

The evolution of a system of vortex filaments, is achieved numerically
by discretizing continuous space curves, $\bs(\xi,t)$, into a large
number of points $\bs_i$ ($i=1,\cdots N$).  The singularity at
$\bs=\br$ is removed in a standard way by considering local and
non-local contributions to the integral.  If $\bs_i$ is the position
of the $i^{\rm th}$ discretization point along the vortex line,
Eq.~(\ref{eq:BS}) becomes \cite{Schwarz1985}

\begin{equation}
\frac{d\bs_i}{dt}=
\frac{\Gamma}{4\pi} \ln \left(\frac{\sqrt{\ell_i \ell_{i+1}}}{a}\right)\bs_i' \otimes \bs_i'' 
-\frac{\Gamma}{4 \pi} \oint_{\cal L^\star} \frac{(\bs_i-\br) }
{\vert \bs_i - \br \vert^3}
\times {\bf d}\br.
\label{eq:BS_sing}
\end{equation}

\noindent
Here $\ell_i$ and $\ell_{i+1}$ are the arclengths of the curve between
points $\bs_{i-1}$ and $\bs_i$ and between $\bs_i$ and $\bs_{i+1}$,
and $\cal L^\star$ is the original vortex configuration without the
section between $\bs_{i-1}$ and $\bs_{i+1}$.

The number of discretization points, $N$, changes with time.  As the
simulation progresses, new discretization points are introduced to
maintain the resolution along the vortex filament.  If the separation
between two points, $\bs_{i}$ and $\bs_{i+1}$, becomes greater than
some threshold $\Delta \xi$, a new point is introduced at position
$\bs_{i'}$ given by
\begin{equation}
\bs_{i'}=\frac{1}{2}(\bs_i+\bs_{i+1})+\left( \sqrt{R^2_{i'}
-\frac{1}{4}\ell_{i+1}^2}-R_{i'} \right)\frac{\bs_{i'}^{''}}{|\bs_{i'}^{''}|},
\end{equation}

\noindent
where $R_{i'}=|\bs_{i'}^{''}|^{-1}$. This approach preserves the
curvature of the vortex filament, whilst ensuring that the separation
of vortex points along a filament lies between $\Delta \xi/2$ and
$\Delta \xi$, where $\Delta \xi=0.05\,$mm

All spatial derivatives are approximated using $4^{\rm th}$ order
finite difference schemes which account for varying mesh sizes along
the vortex filaments.  Time evolution is based on a $3^{\rm rd}$ order
Runge-Kutta scheme, with a time step equals to $\Delta t= 5 \times
10^{-4}\,$s.

It is well known that if quantised vortices become close then they can
reconnect \cite{paoletti2008velocity}.  If two discretization points
become closer to each other than $\Delta \xi$, a numerical algorithm
reconnects the two filaments subject to the criteria that the total
length (as proxy for energy) decreases
\cite{Leadbeater2001}. Self-reconnections (which can arise if a vortex
filament has twisted back on itself) are treated in the same manner.
Other reconnection algorithms exist \cite{KondRecon}, however in a
previous study \cite{BaggaleyRecon} we verified that in uniform
counterflow important properties of the system were robust to the
method of reconnection used.

All calculations are performed in a cuboid of size $D_x \times D_y
\times D_z=4\pi \,\textrm{mm} \times 2 \,\textrm{mm} \times 4\pi/3
\,\textrm{mm}$.  Periodic boundary conditions are used in the $x$ and
$z$ directions, whereas solid boundaries are applied at $y=\pm
D_y/2$. In all simulations the counterflow lies parallel to the
elongated $x$ direction.
The boundaries are assumed to be perfectly smooth, hence no modelling of the pinning of vortices
to any roughness on the boundaries is required. This is justified as the flow velocities considered here are much larger than the modest velocities required to de-pin quantised vortices from the boundary \cite{Schwarz1985}.

\section{Normal Fluid Velocity} \label{section:3}

If we assume that the normal fluid is a laminar flow in the positive
$x$ direction, trapped between solid boundaries at $y=\pm D_y/2$
(assuming no-slip boundaries) then it is appropriate to take the
following simple Poiseuille profile
\begin{equation}
\bv_n= \left(1-\frac{y^2}{h^2}\right) U_c \ \mathbf{e_x},
\end{equation}
where $h=D_y/2$ is the half-width of the channel.  The relative motion
of the two velocity fields, commonly denoted $v_{ns}$, is then simply
given by $v_{ns}=2 U_c/3$, the mean flow velocity along the channel.

In the case of a turbulent channel flow the situation is clearly more
complex. The governing equations, assuming an incompressible normal
fluid ($\boldsymbol{\nabla} \cdot \bv_n = 0$), are the Navier-Stokes
equations

\begin{equation}
\frac{\partial \bv_n}{\partial t} =
- \bnabla p_n
- \frac{1}{2}\left[\bnabla \left(\bv_n \otimes \bv_n\right)
+(\bv_n \cdot \bnabla)\bv_n \right]
+ \nu \bnabla^2\bv_n
+ \mathbf{f}, \qquad
\label{NS}
\end{equation}
where $p_n(\mathbf{x},t)$ is the pressure field (for a fluid with a
constant density $\rho_n=1$), $\nu$ is the kinematic viscosity, and
$\mathbf{f}$ is an externally applied force.

Equations (\ref{NS}) are solved using the code
Incompact3d \footnote{This open source code is now available at
  http://code.google.com/p/incompact3d/}. This code is based on
sixth-order compact finite difference schemes for the spatial
differentiation and a third order Runge-Kutta scheme for the time
integration. To treat the incompressibility condition, a fractional
step method requires solving a Poisson equation. This equation is
fully solved in spectral space, via the use of relevant 3D Fast
Fourier Transforms. The pressure mesh is staggered from the velocity
mesh by half a mesh to avoid spurious pressure oscillations. With the
help of the concept of modified wave number \cite{lele92}, the
divergence-free condition is ensured up to machine accuracy.  Full
details about the code can be found in Laizet \& Lamballais
\cite{laizet&lamballais09}.

\subsection{Numerical parameters}

In order to investigate superfluid turbulence in a channel flow
configuration, a Direct Numerical Simulation (DNS) of a
turbulent channel flow is performed with a Reynolds number of $Re=U_m
h/\nu=2793$ where $U_m$ is the bulk velocity and $h=D_y/2$ is the
half-width of the channel. Using the conventional notation where
superscript $^+$ indicates a scaling based on the friction velocity
$u_\tau$ and the kinematic velocity $\nu$, this Reynolds number leads
to $h^+\approx 178.1$. The numerical domain is based on $n_x\times
n_y\times n_z = 128\times 129\times 84$ mesh nodes to discretize the
computational domain $D_x \times D_y \times D_z=4\pi \,\textrm{mm}
\times 2 \,\textrm{mm} \times 4\pi/3 \,\textrm{mm}$, with $x$, $y$ and
$z$ being the longitudinal, normal to the wall and spanwise directions
respectively.  In the $y$-direction, a stretching is applied in order
to concentrate grid points near the walls at $y=\pm h$ with the first
adjacent point to the wall located at $y^+=1$ (see Laizet \&
Lamballais \cite{laizet&lamballais09} for further details about the
stretching). Periodic boundary conditions are used in $x$ and $z$
directions while no-slip boundary conditions are imposed at the two
walls $y=\pm h$. These boundary conditions are consistent with those
applied to the quantised vortices.

In order to get a fully turbulent state, an initial random
perturbation is superimposed on a Poiseuille profile
\begin{equation}
\bv_n= \left(1-\frac{y^2}{h^2}\right) U_c \ \mathbf{e_x}.
\end{equation}
In order to quickly obtain a fully turbulent state, a rotation in the
$z$ direction is imposed in the early stage of the simulation with a
Rossby number equal to $Ro=U_mh/f=1/18$ where $f$ is the Coriolis
frequency.

\begin{figure}
\begin{center}
\mbox{\includegraphics[width=0.45\textwidth,clip=]{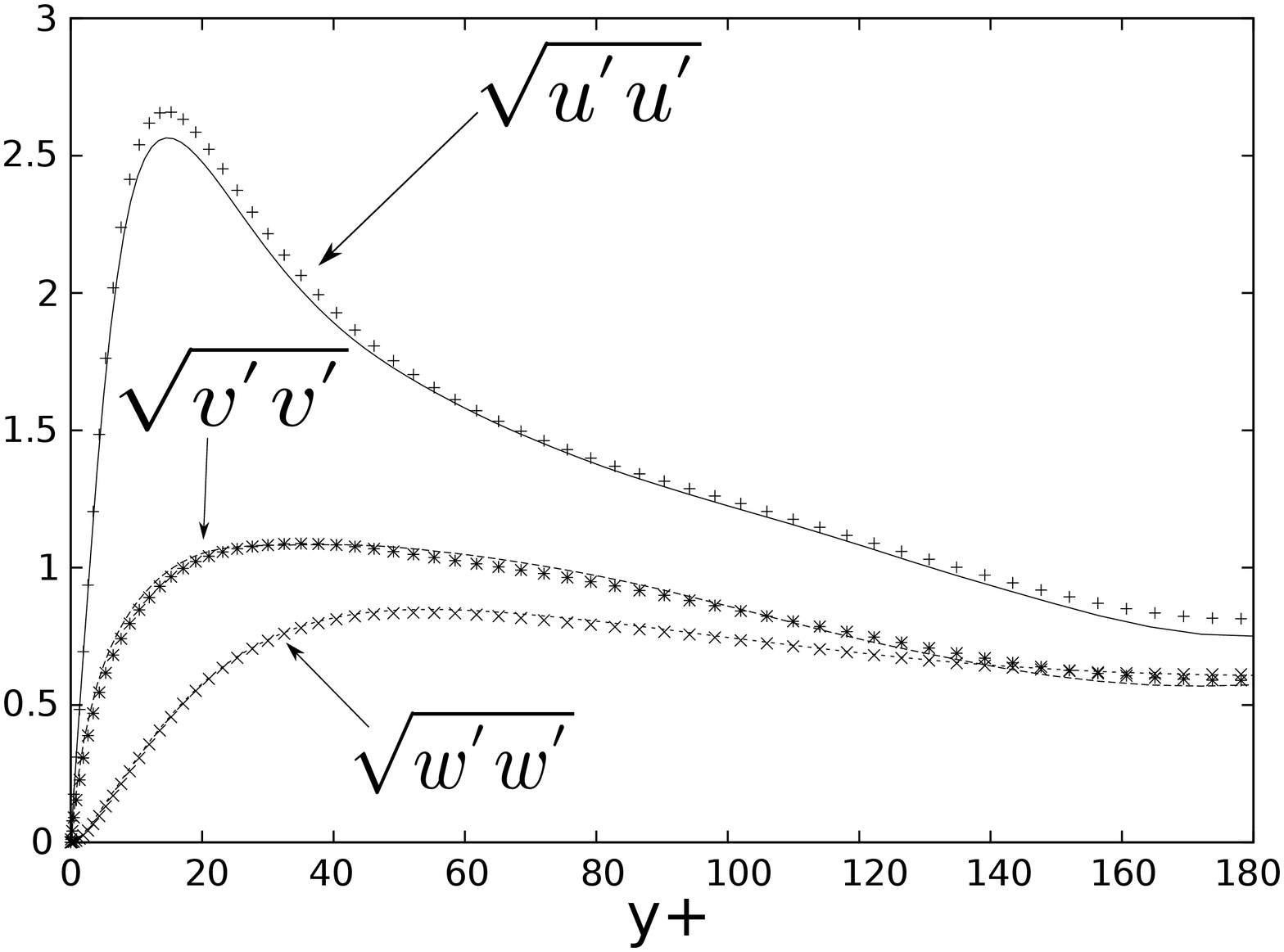}}
\mbox{\includegraphics[width=0.45\textwidth,clip=]{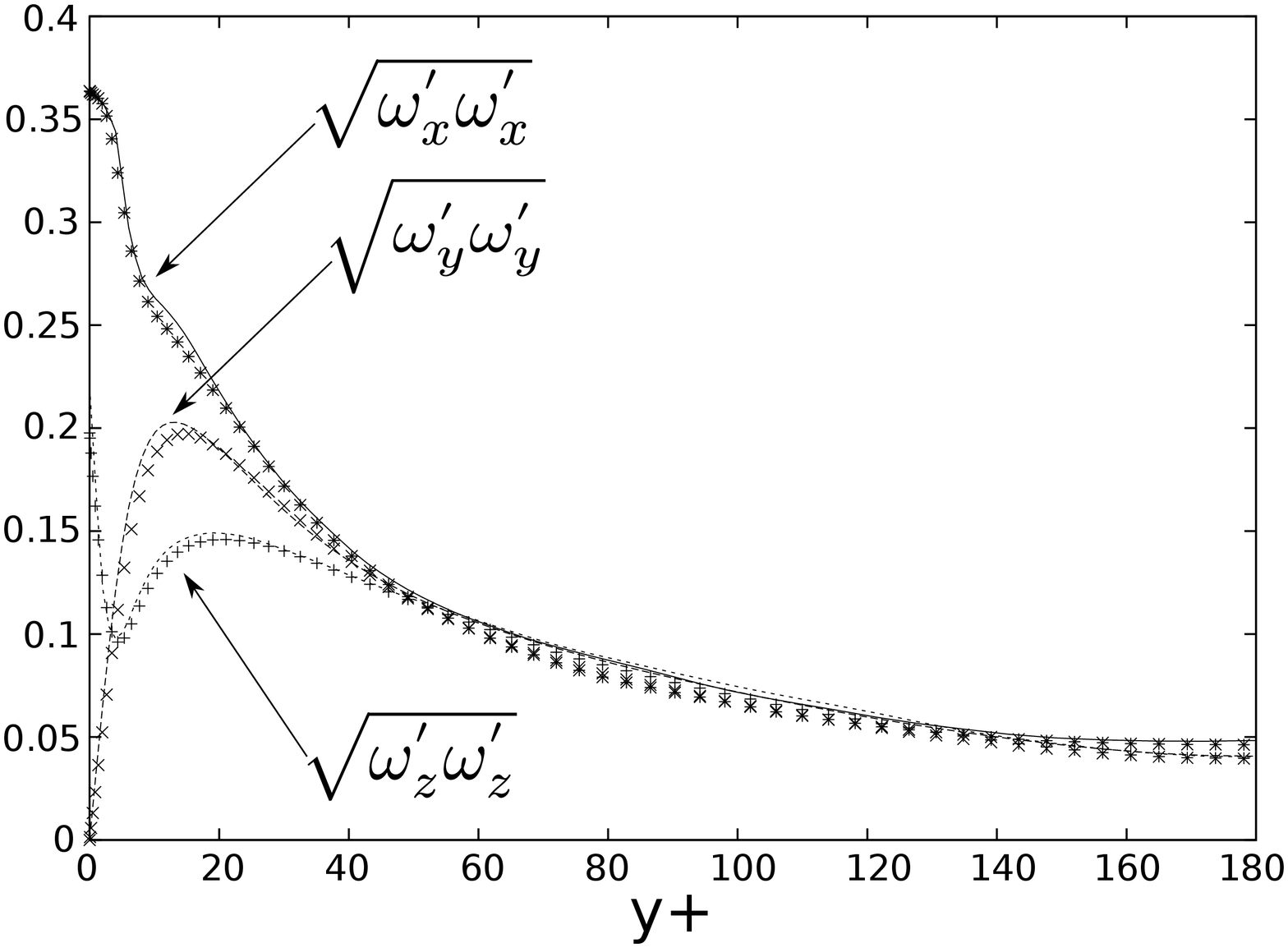}}
\caption[]{Turbulent intensities (left) and r.m.s of the fluctuating
  vorticity components (right) in wall units. Lines: present
  simulation. Symbols: \cite{moseretal99}.}
\label{urms}
\end{center}
\end{figure}

\begin{figure}
\begin{center}
\psfrag{x}{$x$}
\psfrag{y}{$y$}
\mbox{\includegraphics[width=0.85\textwidth,clip=]{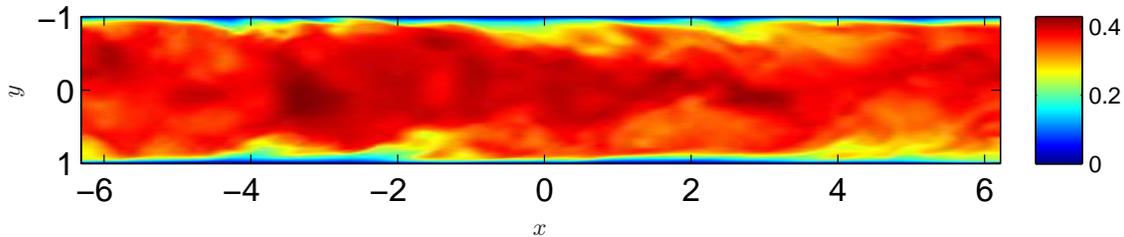}}
\caption[]{Slice of the $x$ component of the turbulent normal fluid in
  the $xy$-plane at $z=0$.}
\label{turbslice}
\end{center}
\end{figure}

The velocity field is then time stepped until a fully developed
turbulent state is achieved; this is then validated against previous
studies.  Profiles of velocity and vorticity r.m.s obtained with our
simulations are presented in figure \ref{urms}.  An excellent
agreement with previous spectral DNS of Moser \etal \cite{moseretal99}
is found. Figure \ref{turbslice} shows the structure of the turbulent
channel flow in the $xy$-plane at $z=0$.

Due to the high computational cost of computing the evolution of the
quantised vortices and the turbulent normal fluid, we follow Kivotides
\cite{Kivotides2006} and use a frozen normal fluid, i.e. a single
snapshot of the turbulent normal fluid is used meaning $\bv_{n}$ is
time independent in both the laminar and turbulent simulations. As
pointed out in \cite{Kivotides2006}, this approach can be loosely
justified at the low normal fluid Reynolds numbers use in this study,
where the turnover time of eddies in the normal fluid is relatively
slow.  In addition in counterflow there is an anisotropy in the
structure of the tangle \cite{Schwarz1988,Adachi2010} leading to the
vortices drifting with respect to the imposed superflow. Hence the
vortices move through the computational domain, and so in the rest
frame of the vortices the flow appears time dependent. In order to
make comparisons between the laminar and turbulent cases we scale
$\bv_n$ such that $\langle \bv_n \cdot \mathbf{e_x} \rangle=v_{ns}=0.34\,$
cm/s, where angled brackets denote spatial averaging.

\section{Results}\label{sec:results}

\begin{figure}
\begin{center}
\psfrag{t}{$t$}
\psfrag{L}{$L$}
\mbox{\includegraphics[width=0.45\textwidth]{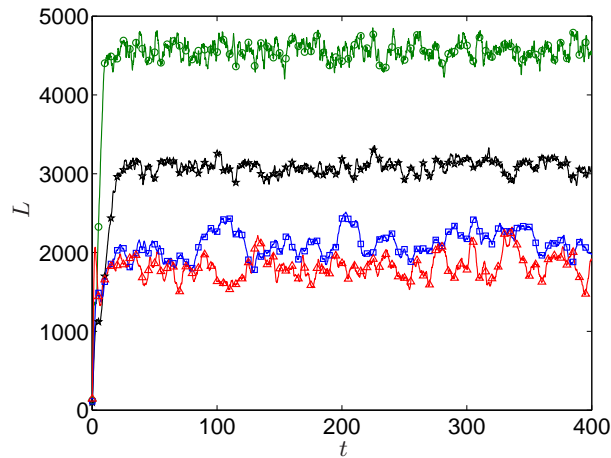}}
\caption[]{The evolution of the vortex line density, $L$ (cm$^{-2}$),
  plotted in time, $t$ (s).  Solid (blue) line plotted with squares -
  laminar $\bv_n$, T=1.6K ; (red) line plotted with triangles -
  laminar $\bv_n$, T=1.9K ; (black) line plotted with stars -
  turbulent $\bv_n$, T=1.6K ; (green) line plotted with circles -
  turbulent $\bv_n$, T=1.9K.}
\label{LineDensity}
\end{center}
\end{figure}

We present the results of four numerical simulations, all initialised
with 100 randomly oriented loops confined to the numerical domain,
with a radius of 0.0597mm. These consist of simulations using a
laminar (Poiseuille) normal fluid and the frozen DNS channel flow, at
two different temperatures T=1.6K ($\alpha=0.097$, $\alpha '=0.0161$ )
and T=1.9K ($\alpha=0.206$, $\alpha '=0.00834$ ). We monitor the
evolution of the vortex line density $L=\Lambda/V$, where $V$ is the
volume of the domain and $\Lambda=\int_{\cal L}d\xi$ is the total
length of the vortex configuration. In all simulations after an
initial transition, where a rapid growth in the vortex line density
$L$ is observed, the system saturates to a non-equilibrium steady
state, as it can be seen in Figure \ref{LineDensity}. Here energy
extracted from the normal fluid (through the Donnelly-Glaberson
instability) is balanced by energy dissipation due to vortex
reconnections \cite{Vinen3,Vinen4}.

At both temperatures we see that the simulations with a turbulent
normal fluid saturate at a much higher vortex line density. Of obvious
interest is how this difference can arise, and if any differences in
the structure of the vortex tangle are apparent. It is also
interesting to note that the vortex line density in the lower
temperature (T=1.6K) laminar simulation is slightly higher than in the
corresponding higher temperature simulation. Initially this may seem
counterintuitive as the timescale for the growth of Kelvin waves by
the Donnelly-Glaberson instability is inversely proportional to
$\alpha$\cite{Tsubota:2004}, and so it would seem reasonable to assume
larger values of $\alpha$ would lead to larger values of $L$. However
this discrepancy can also be explained by examining the structure of
the tangle.

\begin{figure}
\begin{center}
\psfrag{vns}{$v_{ns}$}
\psfrag{RL}{$\sqrt{L}$}
\mbox{\includegraphics[width=0.45\textwidth]{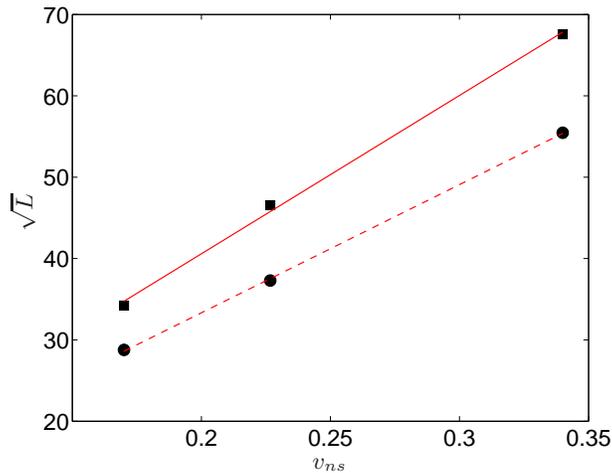}}
\caption[]{The square root of the vortex line density $L$ ($\rm cm^{-2}$) vs $v_{ns}$ ($\rm cm/s$) for simulations with a turbulent normal fluid at T=1.6K (circles) and T=1.9K (squares). Lines of best fit are used to compute the scaling coefficient $\gamma$ (s/cm$^{2}$); we find  $\gamma=195\,$s/cm$^2$ (1.9K) and $\gamma=157\,$s/cm$^2$ (1.6K).}
\label{gamma}
\end{center}
\end{figure}

\subsection{Vortex line density versus counterflow velocity}
In his seminal works Vinen \cite{Vinen3,Vinen4} introduced a phenomenological
model for the evolution of the vortex line density $L(t)$ in a homogeneous vortex tangle.  
In a steady state regime ($dL/dt=0$) Vinen's equation yields the following scaling law
\begin{equation}\label{eq:gammarelation}
\sqrt{L} = \gamma(T) (v_{ns}-v_0),
\end{equation}  
where $v_0$ is an additional fitting parameter, and $\gamma$ is a temperature dependent parameter.  The measurement of $\gamma$ and the verification of this relationship is well established in the literature, across both experimental \cite{Tough1982,Babuin12} and numerical \cite{Adachi2010,Baggaley_tree2012} studies.

However at a fixed temperature the value of $\gamma$ was found to dramatically increase when the applied heat flux (i.e. the counterflow velocity $v_{ns}$) became larger than some critical value. This led Tough \cite{Tough1982} to categorised two turbulent regimes TI and TII. Using linear stability analysis Melotte \& Barenghi \cite{Melotte1998}  suggested that transition to turbulence in the normal fluid was the underlying reason for the transition from the TI to TII regime. With the caveat of the steady imposed flow used in this study, we are in a position to shed some light on this problem, by computing $\gamma$ for the turbulent normal fluid, and comparing the results to experimental \cite{Tough1982,Babuin12} and numerical \cite{Adachi2010} results.

We perform further simulations using the turbulent (DNS) channel flow and rescaling the counterflow velocity to lower values of $v_{ns}=0.23\,$ cm/s and $0.17\,$cm/s, at both T=1.6K and 1.9K. 
Figure \ref{gamma} shows the familiar linear scaling of the square root of the vortex line density $L$ with the imposed counterflow velocity. The gradient of the line of best (least squares) fit to the data gives $\gamma=195\,$s/cm$^2$ (1.9K) and $\gamma=157\,$s/cm$^2$ (1.6K). These are significantly larger than the values quoted in Adachi \etal \cite{Adachi2010}, where a laminar (constant) counterflow was imposed, and are comparable to experimental values \cite{Babuin12}, in the TII regime. Hence turbulence in the normal fluid is certainly capable of leading to larger values of $\gamma$, lending the first firm numerical evidence of a transition to turbulence in the TI to TII transition. 

\subsection{Structure of the tangle}\label{subset:structure}
\begin{figure}
\begin{center}
\mbox{\includegraphics[width=0.25\textwidth]{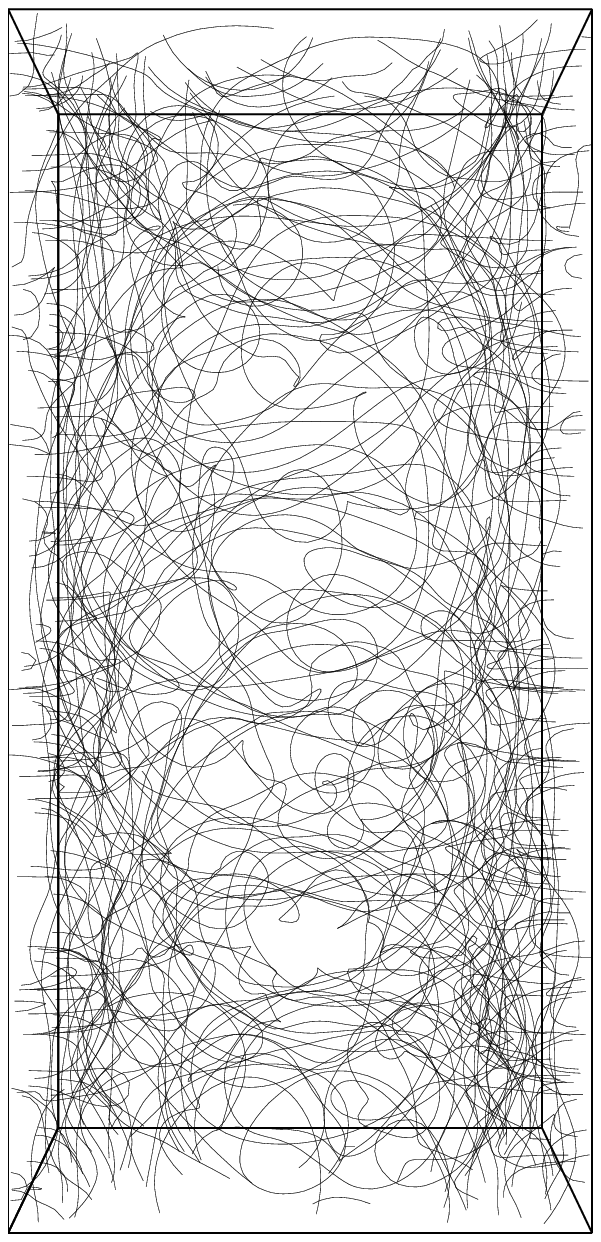}}
\hspace{8mm}
\mbox{\includegraphics[width=0.25\textwidth]{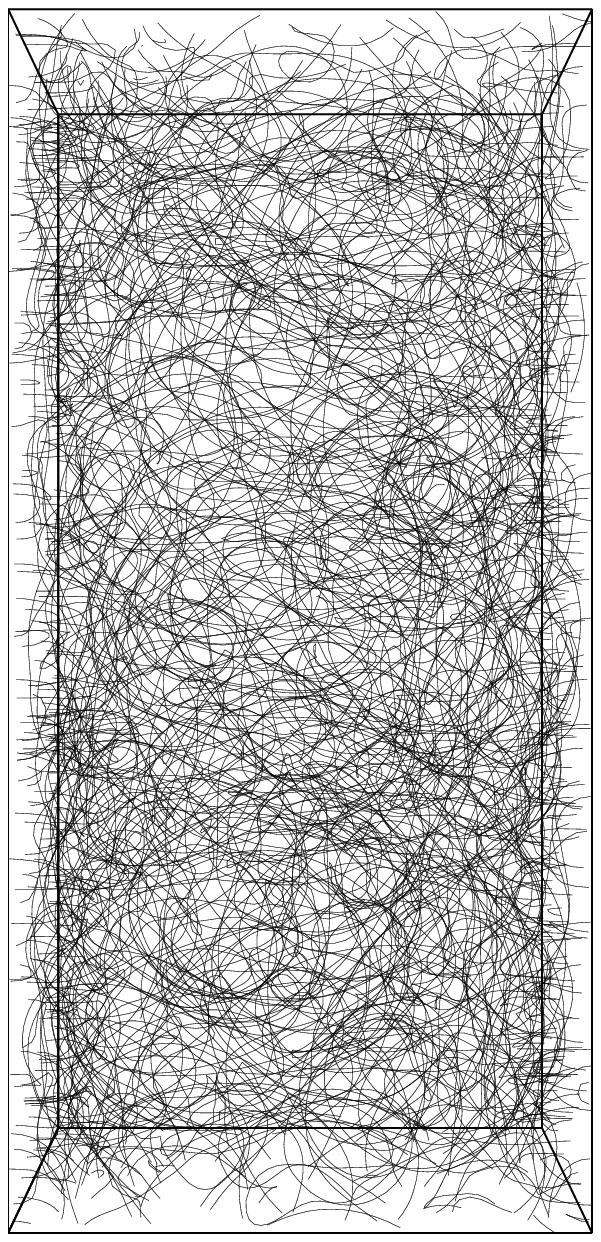}}
\caption[]{Snapshots of the vortex configuration in the laminar (left)
  and turbulent (right) simulations, at T=1.9K, and $t=400$s, plotted
  in the $yz$-plane. Notice the much higher vortex density in the
  turbulent case, as well as a more homogeneous distribution of
  vortices.}
\label{sidetangles}
\end{center}
\end{figure}

\begin{figure}
\begin{center}
\psfrag{L}[b]{$L'/L$}
\psfrag{y}{$y$}
\mbox{\includegraphics[width=0.45\textwidth]{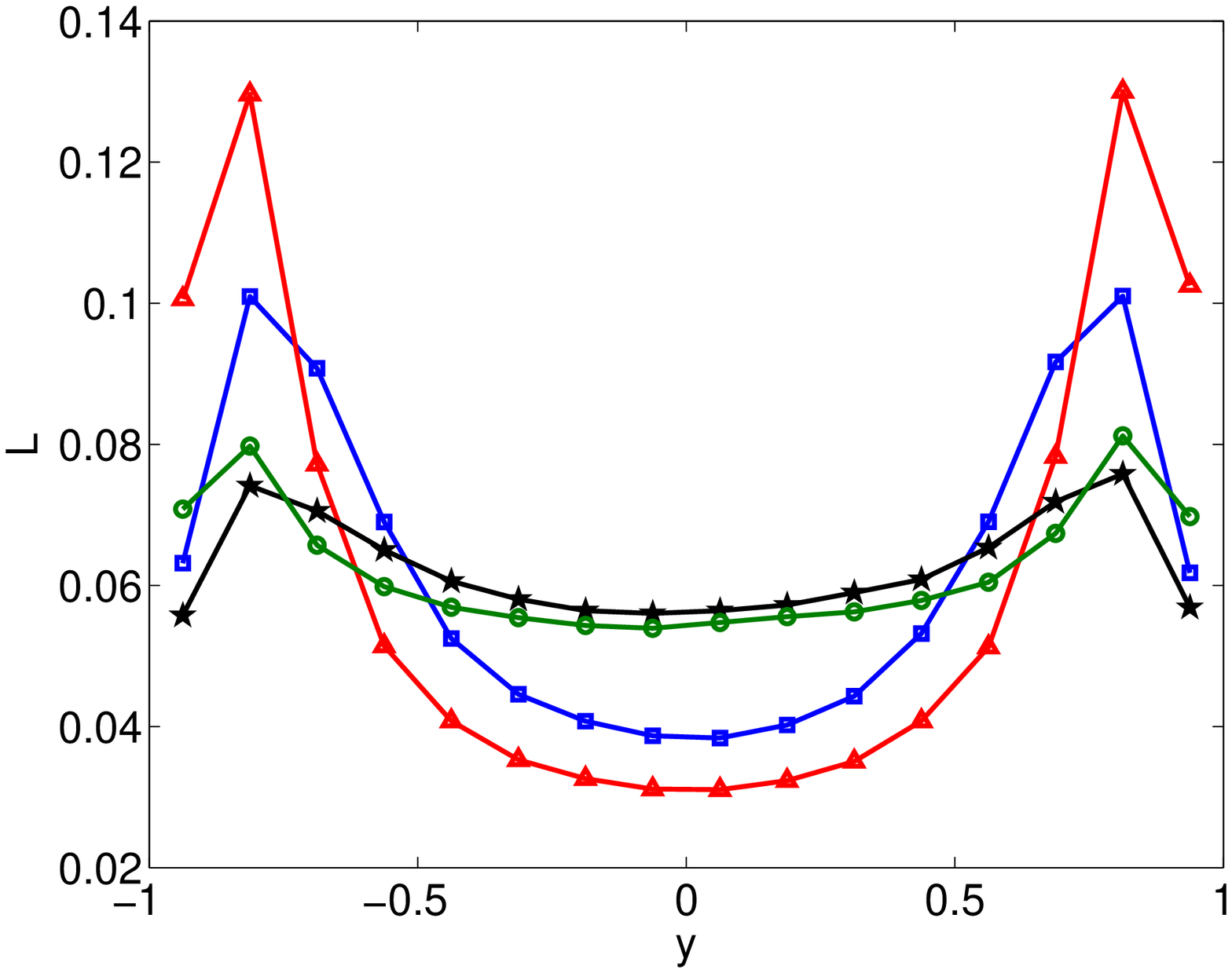}}
\hspace{4mm}
\psfrag{curv}[b]{$\zeta '$}
\mbox{\includegraphics[width=0.45\textwidth]{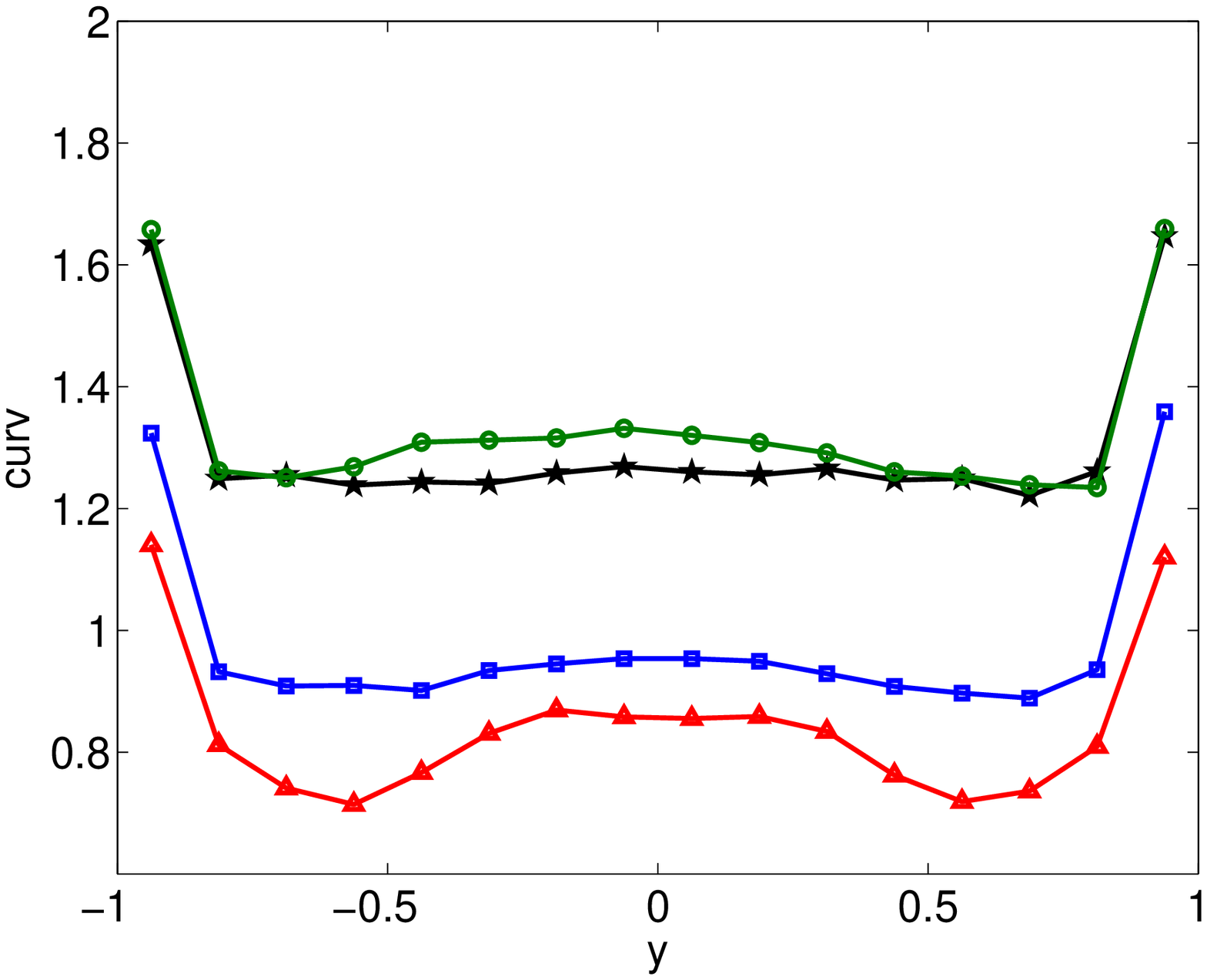}}
\caption[]{Course grained statistics of the tangle plotted as function
  of the wall normal direction, $y$. Line style (colour) is as in
  Fig.\ref{LineDensity}. The left panel displays the spatial
  distribution of the vortex line density $L'(y)$, normalised by the
  total vortex line density $L$ ; the right panel displays $\zeta'(y)$
  the spatial distribution of the curvature $\kappa '(y)$, scaled by
  the inter vortex spacing $\ell '(y)$.}
\label{CG1}
\end{center}
\end{figure}

Figure \ref{sidetangles} displays snapshots of the vortex tangle when
driven by a laminar (left) and turbulent normal fluid (right) in the
higher temperature simulations (T=1.9K), plotted in the
$yz$-plane. Clear differences in the spatial homogeneity of the
tangles are visible. In order to quantify these differences we perform
a course graining of the statistics of the tangle in time. A regular
$16^3$ Cartesian mesh is defined in the computational volume. Within
each discretization volume in this mesh, course grained information
about the tangle is computed, including the total vortex line density
within each sub-volume, the mean curvature and the mean velocity. We
average this course grained mesh in time, over 3500 realisations,
computed at 0.1 second intervals; subsequently we average over the two
periodic directions $x$ and $z$ to find average measures as a function
of $y$, i.e. across the channel. More formally we can define these
quantities as,
\begin{equation} \label{CGL}
L'(y)=\left \langle \dfrac{1}{\cal V'}\int_{\cal L'} d\xi \right \rangle
\end{equation}
\begin{equation}\label{CGC}
 \kappa '(y)=\left \langle \int_{\cal L'} |\bs(\xi)''|d\xi \middle /
 \int_{\cal L'} d\xi \right \rangle
\end{equation}
\begin{equation}\label{CGV}
 v_{x,y,z} '(y)=\left \langle \int_{\cal L'} \dfrac{ds_{x,y,z}}{dt}
 d\xi \middle / \int_{\cal L'} d\xi \right \rangle
\end{equation}

where $\cal L'$ denotes the vortex configuration within the reduced
volume, $\cal V'$, each element of the Cartesian mesh encloses; angled brackets
denote the temporal and spacial averaging procedure defined
above. Clearly from Eq.~(\ref{CGL}) one can define an equivalent inter
vortex spacing $\ell '(y)=1/\sqrt{L'}$. Figure \ref{CG1} (left) shows
the course grained line density plotted along the channel. It is clear
that with a laminar normal fluid the vortex line density is
concentrated closer to the wall, in contrast to the turbulent
simulations where the distribution of vortex density is more
homogeneous, visible in the snapshots of the vortex tangles presented
in Figure \ref{sidetangles}.

As we have discussed the steady state value of the vortex line density
is a complex interplay between the energy transferred via the
Donnelly-Glaberson (DG) instability and the energy dissipated
due to vortex reconnections. The DG instability is the amplification
of Kelvin waves (helical displacements of the vortex cores) by the
component of the counterflow velocity parallel to the vortex, hence
seed disturbances, wiggles along the vortices, are key for converting
normal fluid energy to vortex length. This is balanced by the density
of vortices which controls the rate of reconnection
\cite{reconscaling}. Therefore, a useful quantity to study is the
dimensionless variable $\zeta=\kappa / \sqrt{L} = \kappa \ell $, with
larger values providing an explanation for larger global vortex line
densities. A course grained form, $\zeta '$, of this quantity can
readily be computed using Eqns.~(\ref{CGL}) \& (\ref{CGC}). This
quantity is displayed in Figure \ref{CG1} (right). It can be
  seen that we find larger values corresponding to larger values of
the global vortex line density $L$, plotted in Figure
\ref{LineDensity}. Therefore, one of the effects of turbulence in the
normal fluid is to create higher curvature along the vortices, which
promote the DG instability and allow a greater vortex line density to
be reached. From inspection of Figure \ref{CG1} (right) it is now
perhaps clear why, in the laminar simulations, a marginally higher
steady-state vortex line density is reached in the lower temperature
simulation. As Kelvin waves are not damped as strongly at lower
temperatures (due to reduced friction), vortex lines tend to have
larger curvature, and so at least in these simulations a perhaps
unexpected result is found.

\begin{figure}
\begin{center}
\psfrag{y}{$y$} \psfrag{vxvyvz}[b]{$ v'_x $,\,$ v'_y $,\,$ v'_z $}
\mbox{\includegraphics[width=0.5\textwidth]{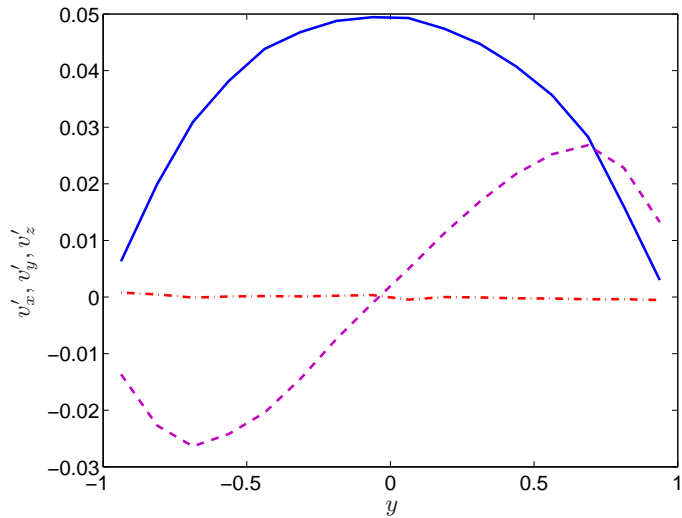}}
\caption[]{Course grained components of the drift velocity of the
  vortices $\bv_i '(y)$, as defined in Eq.~\ref{CGV}. The solid (blue)
  line represents $ v'_x $; (purple) dot-dashed $ v'_y $; (red)
  dot-dashed $ v'_z $. These results are computed from the higher
  temperature (T=1.9K) turbulent normal fluid simulation.}
\label{CG2}
\end{center}
\end{figure}

\begin{figure}
\begin{center}
\psfrag{y}[t]{$y$}
\psfrag{vy}[b]{$v_x'/\overline{ |v_x'|}$}
\mbox{\includegraphics[width=0.31\textwidth]{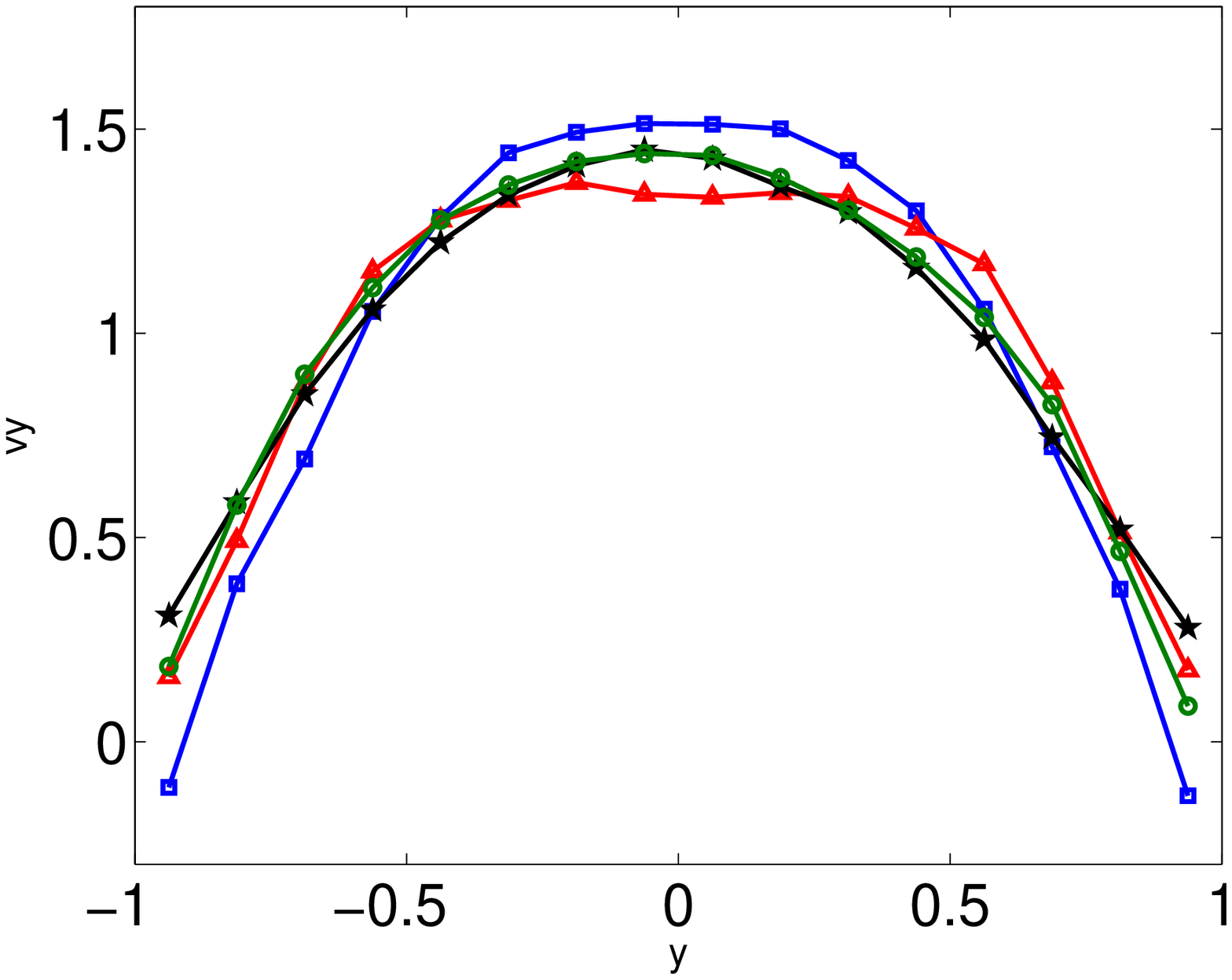}}
\hspace{1mm}
\psfrag{vy}[b]{$v_y'/\overline{ |v_y'|}$}
\psfrag{vz}[b]{$v_z'/\overline{ |v_z'|}$}
\mbox{\includegraphics[width=0.31\textwidth]{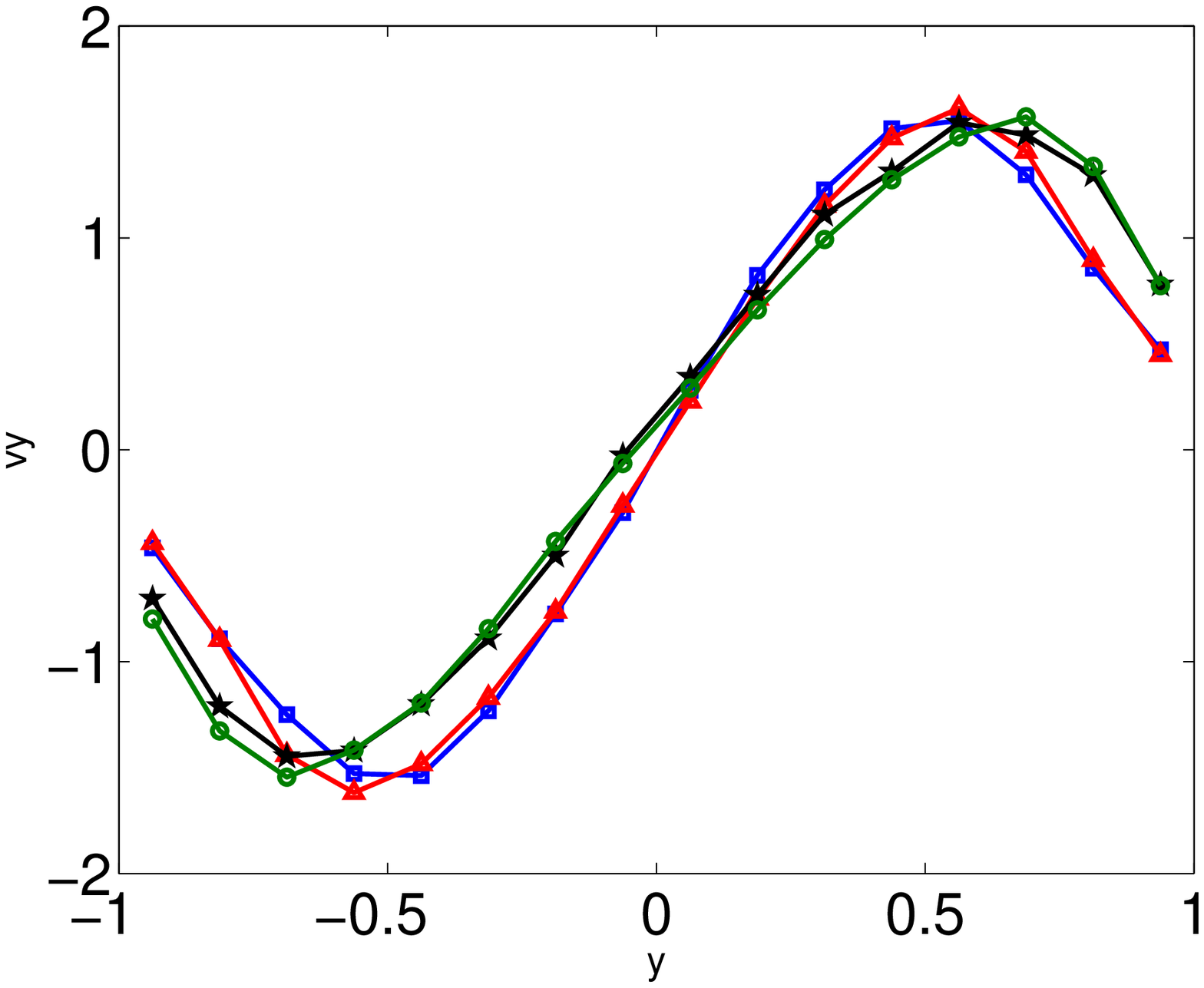}}
\hspace{1mm}
\mbox{\includegraphics[width=0.31\textwidth]{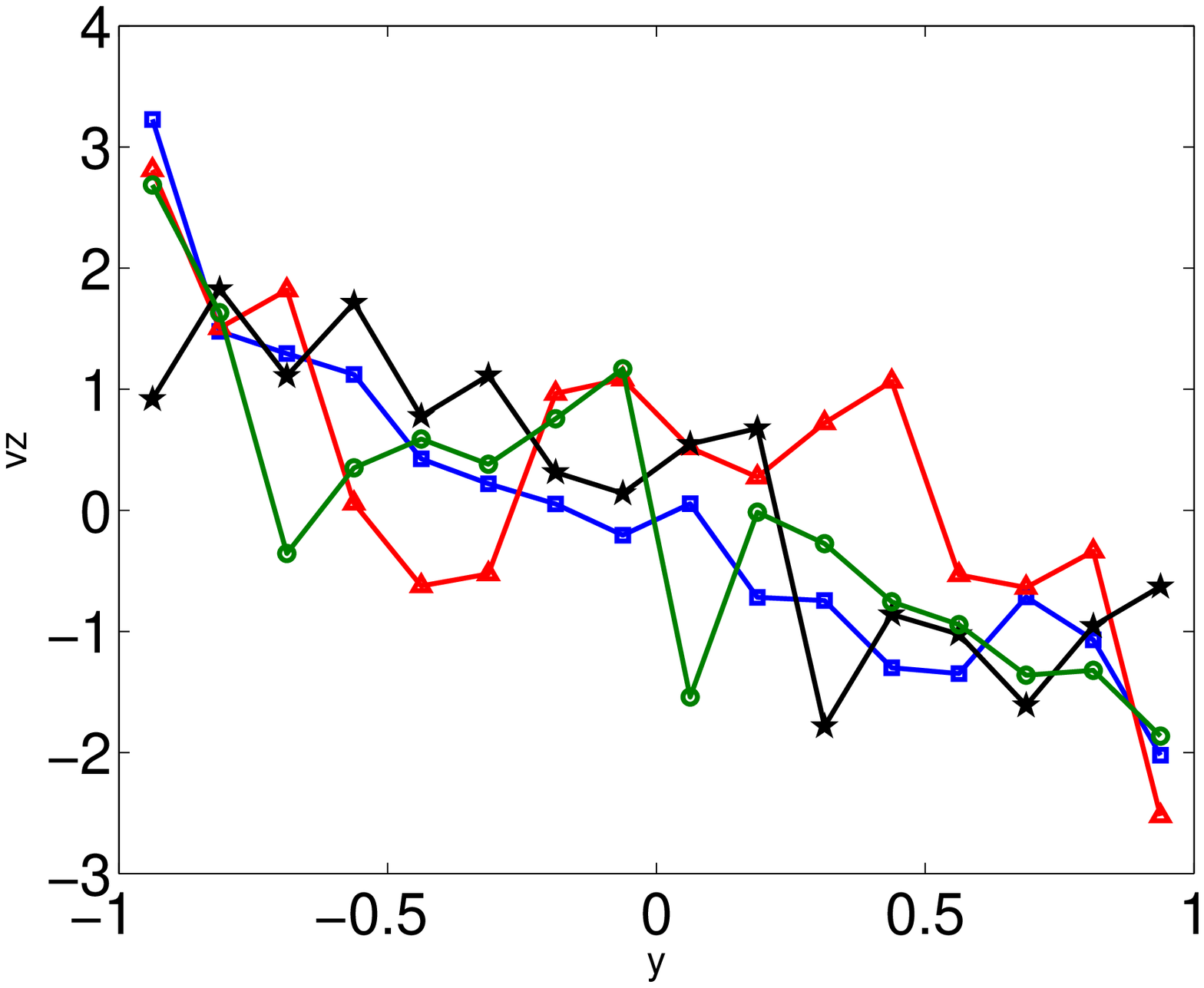}}
\caption[]{Normalised course grained components of the drift velocity
  of the vortices $\bv=d{\bf s}/{dt}$, as defined in Eq.~\ref{CGV}
  plotted for the four simulations (line style (colour) is as in
  Fig.\ref{LineDensity}), the overbear denotes averaging in the $y$
  direction. }
\label{CG3}
\end{center}
\end{figure}

Lending further weight to the importance of the curvature of the
vortices is the fact that the profile of the average drift velocities
of the vortices (see Equation \ref{CGV}) is very similar in all
simulations. Figure \ref{CG2} shows unscaled values of each of the
components of $\bv'$ computed from the higher temperature (T=1.9K)
turbulent normal fluid simulation, plotted as a function of
$y$. Results from the other simulations are comparable with mean
values of the $x$ and $y$ components of the vortices' velocity being
comparable (in terms of magnitude) and the magnitude of the mean value
of the $z$ component being orders of magnitude smaller. In order to
compare between simulations we normalise each component of the
course-grained velocity by the mean of the absolute value of the same
quantity, plotting $v_i(y)'/\overline{ |v_i(y)'|}$ (where $i=x,y,z$
and the overline denotes averaging in the $y$ direction) in Figure
\ref{CG3}. In all simulations we see the vortices in the centre of the
channel move in the direction of the normal fluid, while being pushed
to the walls of the channel by the $y$ component of the mutual
friction. Slight differences in the mean $y$ component of the velocity
between the turbulent and laminar simulations are visible, most likely
due to different profiles of the mean streamwise velocity.

\begin{figure}
\begin{center}
\mbox{\includegraphics[width=0.3\textwidth]{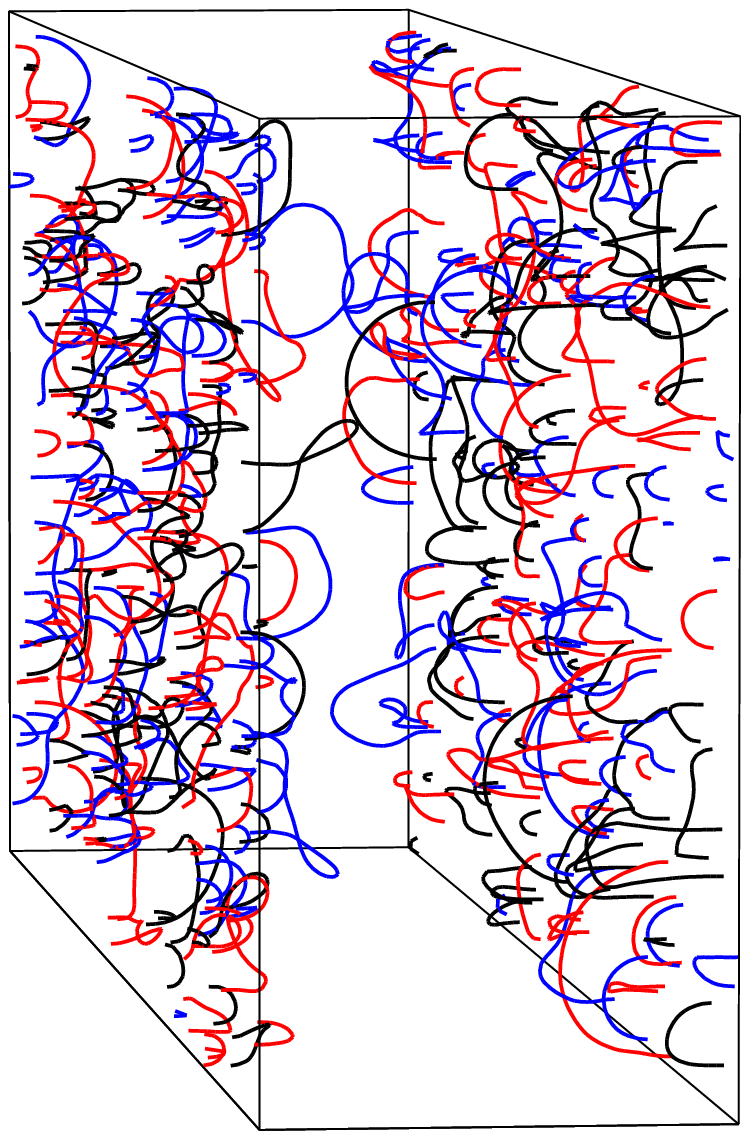}}
\mbox{\includegraphics[width=0.8\textwidth]{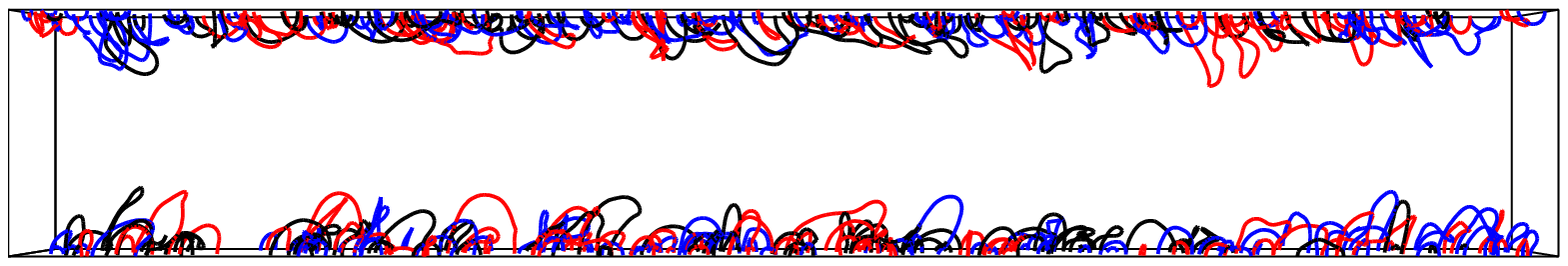}}
\caption[]{The structure of vortices near the sold
  boundaries. Vortices which start and end on the same solid boundary
  are plotted for the simulation with a turbulent normal fluid at
  T=1.9K. Red lines are at $t=100$s, black lines at $t=250$s, and blue
  lines at $t=400$s. The lower panel, plotted in the $xy$-plane shows
  that the vortex loops have a tendency to be angled 'downstream' (the
  normal fluid velocity flowing left to right here).}
\label{hairpin}
\end{center}
\end{figure}

 In all simulations it is interesting to note that although the mean
 $z$ component of the drift velocity is very small, it is not random
 as one might initially expect. This is probably due to the fact that
 small loops moving in the direction of the normal fluid increase in
 length (due to the DG instability) before reconnecting with the walls
 of the container leaving small hairpin vortices (see Figure
 \ref{hairpin}) which are oppositely oriented on each side of the
 channel. If these vortices (initially moving in the $x$ direction)
 are preferentially twisted to lie in the $xy$-plane then on each side
 of the channel they would travel in opposite directions, as seems to
 be indicated by Figure \ref{CG3}.

This near wall `boundary layer' is also particularly interesting, if
one examines the structure of the vortices here. Visible in the scaled
course grained curvature plot (see Figure \ref{CG1}) is the noticeably
higher curvature region close to the solid boundary. To probe the
structure we plot all vortex lines which start and end on the
\emph{same} solid boundary. These vortices (at three different times)
are displayed in Figure \ref{hairpin}, for the turbulent normal fluid
simulations at T=1.9K. Results for the other simulations are
consistent, with a varying number of loops depending on the density at
the boundary. At least qualitatively their appearance is reminiscent
of hairpin vortex structures often identified in turbulent boundary
layers in classical fluids \cite{Robinson1991}.

\begin{figure}
\begin{center}
\psfrag{LxLyLz}{$\Lambda_i$}
\psfrag{1}{$x$}
\psfrag{2}{$y$}
\psfrag{3}{$z$}
\mbox{\includegraphics[width=0.45\textwidth]{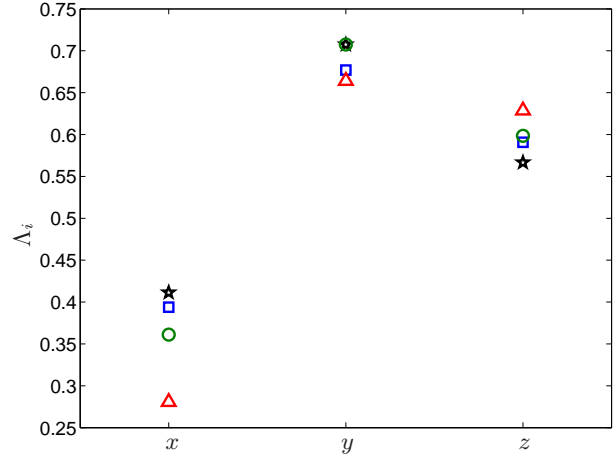}}
\caption[]{Vortex length parallel to each of the Cartesian directions
  $\Lambda_i$, see Eq.~(\ref{Eq:isotropy}), where $i=x,y,z$.  (blue)
  Squares - laminar $\bv_n$, T=1.6K ; (red) triangles - laminar
  $\bv_n$, T=1.9K ; (black) stars - turbulent $\bv_n$, T=1.6K ;
  (green) circles - turbulent $\bv_n$, T=1.9K.}
\label{anisotropy}
\end{center}
\end{figure}

 We also inspect the isotropy of the vortices. For each snapshot of
the vortex tangles, in the saturated quasi-steady state, we compute
the total vortex length parallel to each of the Cartesian directions,
normalised by the total vortex line length and then averaged in time.
More specifically we denote
\begin{equation}\label{Eq:isotropy}
\Lambda_i = \left \langle \int_{\cal L} \bs' \cdot \mathbf{e}_i \,
d\xi \middle / \int_{\cal L} \, d\xi \right \rangle,
\end{equation}
 where $i=x,y,z$ and the angled brackets denote averaging in
 time. Figure \ref{anisotropy} displays the variation of these
 statistics across the four simulations in the study; the results are
 constant with previous studies \cite{Adachi2010}. In all simulations
 the tangles are highly anisotropic with vortices lying in planes
 perpendicular to the counterflow. In addition the vortex
 configuration is more anisotropic at higher temperatures. Turbulence
 in the normal fluid leads to a marginal increase in isotropy of the
 tangle, which would probably be more pronounced with increasing
 Reynolds number.

\subsection{Polarization of the tangle} \label{subsec:fluc}

\begin{figure}
\begin{center}
\psfrag{ss}{$P$}
\psfrag{y}{$|y|$}
\mbox{\includegraphics[width=0.45\textwidth]{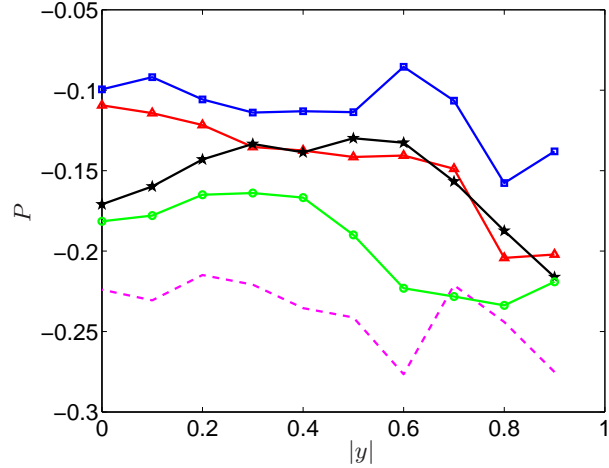}}
\caption[]{The polarization of the vortex tangle computed from Eq.~(\ref{Lvovgamma2}), using discs in the $xz$-plane, at a fixed distance from the channel walls ($|y|=1$). The four curves marked with symbols are computed from the full vortex configurations (e.g. Fig.~\ref{sidetangles}) of the four simulations discussed in the study (line style (colour) is as in
  Fig.\ref{LineDensity}). The dashed (magenta) line is from the DNS simulation (T=1.9K), but only accounting for points which lie within $\ell/2$ of another vortex segment.}
\label{polarise}
\end{center}
\end{figure}

Following \cite{Lvov2007,Baggaley_bundle} we further examine the structure of the vortex tangle through a measure of the vortex polarization. L'vov \etal \cite{Lvov2007}, quantified the polarization of the tangle  by defining the circulation, $\Gamma(R)$ over a contour of a two--dimensional disc of radius $R$.
Vortices  intersect this disc, in both the positive and negative direction, and $\Gamma(R)$ is given by
\begin{equation}\label{Lvovgamma}
 \Gamma(R)=\kappa(N_{+}-N_{-}).
\end{equation}
Assuming the following scaling
\begin{equation}\label{Lvovgamma2}
 \langle\Gamma^2\rangle=\kappa^2(R/\ell)^\sigma,
\end{equation}
then L'vov \etal \cite{Lvov2007} showed that an unpolarized tangle 
has $\sigma=2$, and a tangle with the Kolmogorov ($k^{-5/3}$) spectrum leads to $\sigma=8/3$. From the exponent $\sigma$, the polarisation of the tangle can then be defined as $P=\sigma/2-1$.  The value of $P$ depends on the spectral slope of the flow induced by the tangle, and so is an interesting and informative statistics to compute. 
We define a set of random discs in the $xz$-plane with radius $\ell \le R \le h$, at a fixed distances from the channel walls. This allows us to examine the polarisation as a function of the wall normal direction, following the analysis performed in \S \ref{subset:structure}. We average our results over 10000 discs, at a given radius, for 3500 snapshots of the tangle in the steady-state regime; our results are displayed in Fig.~\ref{polarise}. In all simulations we find that the tangle displays a small negative polarisation, i.e. there is a tendency for anti-parallel vortices to lie close to each other. It is worth commenting that L'vov \etal \cite{Lvov2007} state that the tangle in counterflow will be unpolarized ($P=0$). One possible cause of negative polarisation are vortex reconnections, which dominate the dynamics of counterflow \cite{Sherwin2012}. During a vortex reconnection the induced motion of the vortices tends to reduce any polarization, by turning the vortices as they approach. After reconnection, high curvature is created where vortices lie antiparallel to one another. Reconnections are more likely where the vortex line density is highest, which is closest to the walls, see Fig.~\ref{CG1}. As we would expect, based on the above reasoning,  in all simulations the largest magnitude of $P$ is also found closest to the walls.

We can probe this further, for the high temperature (T=1.9K) simulation performed with the turbulent normal fluid we repeat the above analysis, but only including points which lie within $\ell/2$ of another vortex segment. Our reasoning is that these are likely places where a reconnection has taken place or is about to occur. We find this leads to the largest value of negative polarisation, with $P \approx -0.25$ across the channel. Hence the seemingly random tangles generated by counterflow are not completely random, and exhibit some organisation, one plausible cause is vortex reconnection.

\subsection{Fluctuations of the vortex line density} \label{subsec:fluc}

Studying the spectral properties of fluctuating quantities has proven
to be a powerful technique in the analysis of turbulent flows. Hence,
it is not surprising that a number of experimental studies (see Tough
\cite{Tough1982} for an overview) have analysed the power spectral
density (PSD) of various quantities, most notably the fluctuations of
$L$, in thermal counterflow. However as yet there is no consensus
opinion on the form of the PSD, as a function of the frequency
$f$. From heron we shall refer to PSD($f$) as the power spectral
density of the fluctuating component of the vortex line density,
$\delta L$, where $L=L_0+\delta L$ is decomposed into a mean, $L_0$,
and fluctuating part. An early study by Hoch \etal \cite{Hoch1975}
observed a PSD which scaled as $f^{-1}$ at low frequencies, and
$f^{-3}$ at high frequencies. This contrasts with Ostermeier \etal
\cite{Ostermeier1980} and Barenghi \etal \cite{Barenghifluc82} who
report a flat spectrum at low frequencies, but again with a spectrum
consistent with $f^{-3}$ at higher frequencies (specifically
\cite{Barenghifluc82} report $f^{-2.7 \pm 0.4}$). It is worth noting
that if one considers the phenomenological model of Vinen
\cite{Vinen3,Vinen4}, then a PSD, flat in the low frequency region,
and decaying as $f^{-2}$ is expected \cite{Barenghifluc82}.

\begin{figure}
\begin{center}
\psfrag{PSD}{PSD}
\psfrag{f}{$f$}
\mbox{\includegraphics[width=0.45\textwidth,clip=]{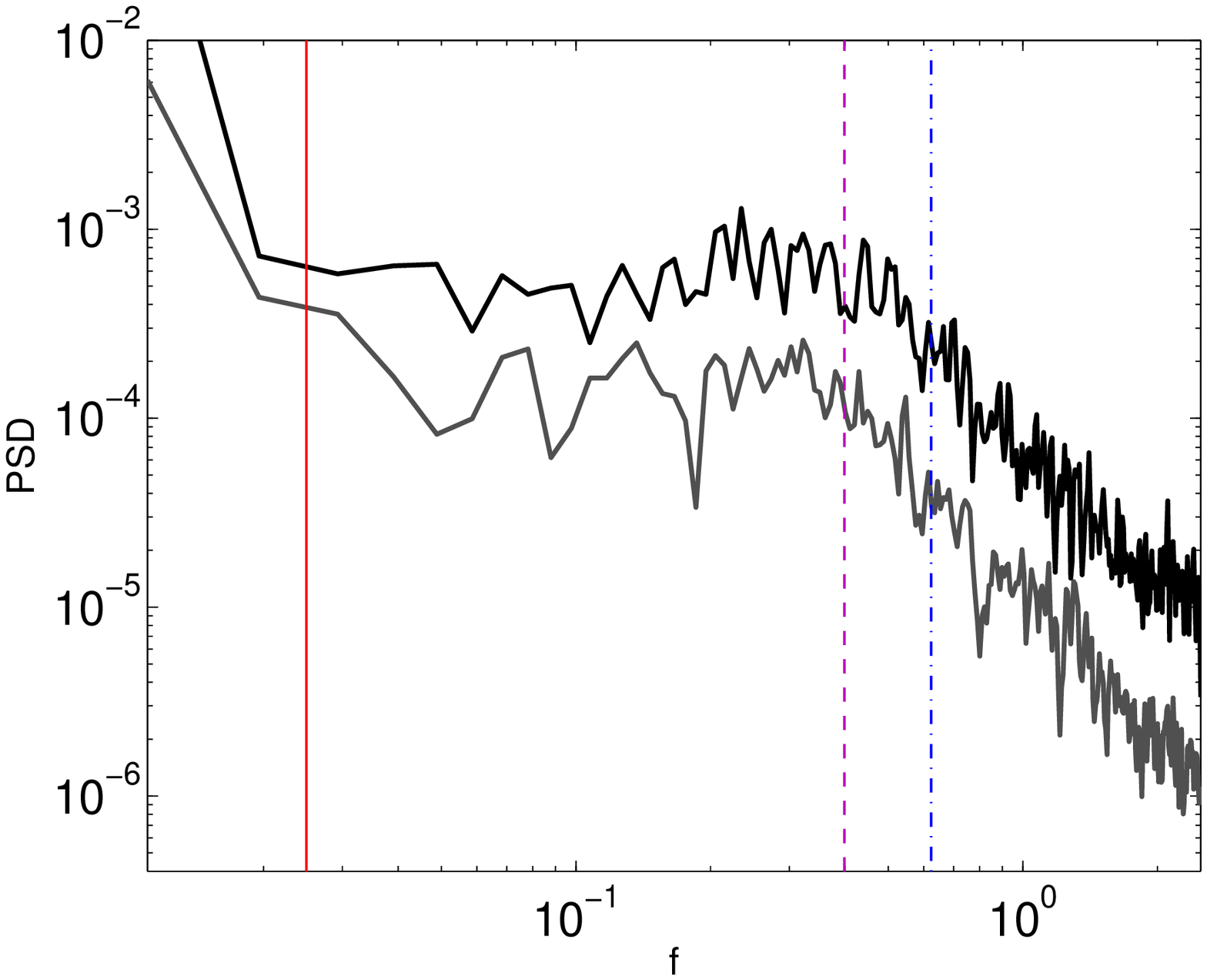}}
\psfrag{fit}{$f^{-3}$}
\mbox{\includegraphics[width=0.45\textwidth,clip=]{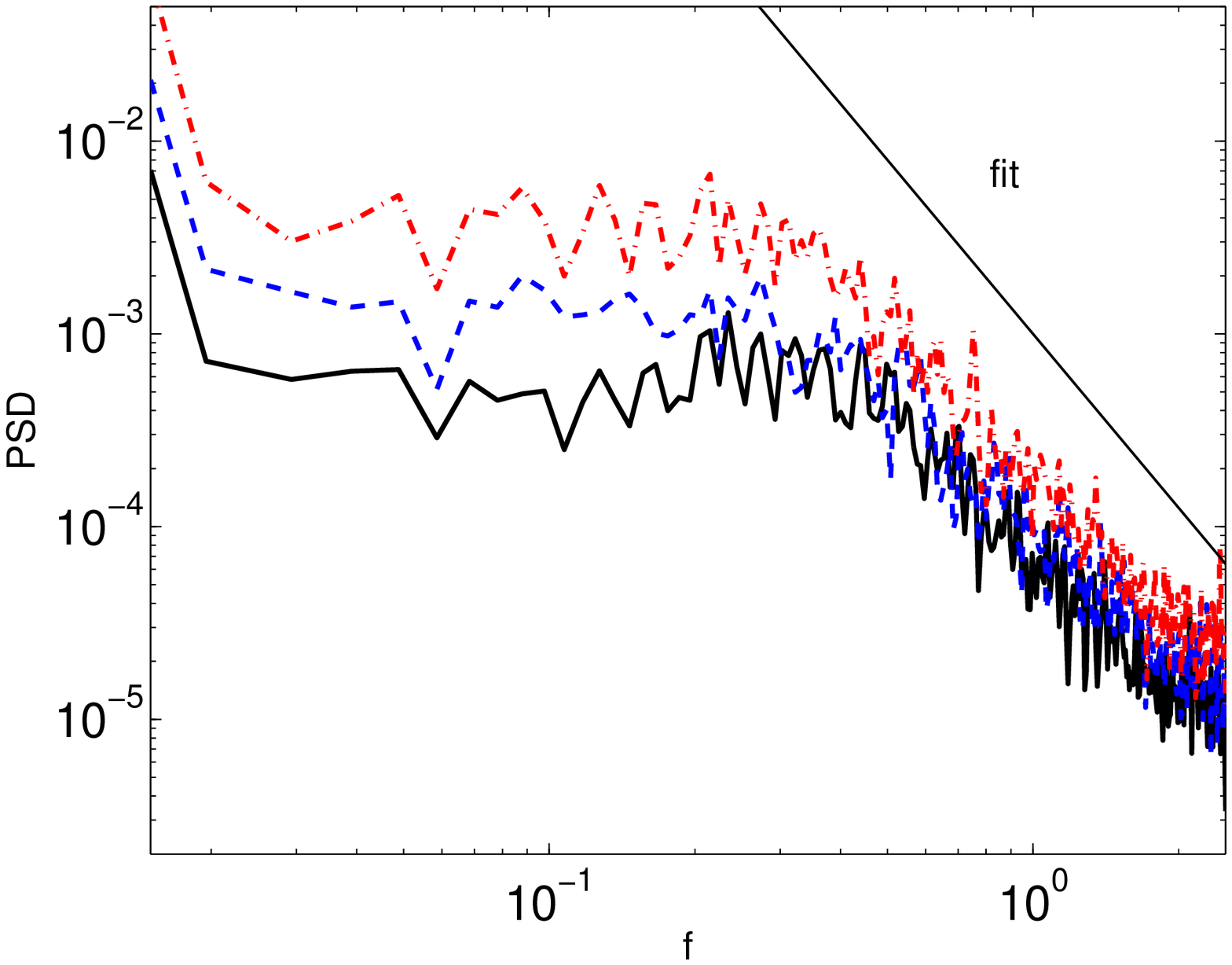}}
\caption[]{Power spectral densities of fluctuations of the vortex line
  density within a volume of the computational domain (arbitrary
  units, scaled for display purposes) vs. $f$ (s$^{-1}$) computed
  during the saturated regime ($50 \, {\rm s} \le t \le 400 \, {\rm
    s}$) . (Left) For a comparison between the laminar (lower grey
  line) and turbulent (upper black line) simulations (T=1.9~K)
  computed in a 1 mm cube at $x=y=z=0$. The solid vertical (red) line
  corresponds to the frequency $\Gamma/D_y^2$, the two dashed vertical
  lines on the right correspond to the frequencies $\Gamma/\ell^2$,
  dashed (purple) for the laminar simulation, dot-dashed (blue) for
  the turbulent simulation. (Right) PSDs of density fluctuations in
  the turbulent simulation (T=1.9~K) computed from different volumes
  in the channel: black solid line - 1mm cube at $x=y=z=0$; dashed
  (blue) line - 1mm cube at $x=z=0, \, y=0.5$mm; dot-dashed (red) line
  - 2mm cube at $x=y=z=0$.}
\label{LPSD}
\end{center}
\end{figure}

Our goal here is not to perform a detailed test of theory, or debate
the validity of experimental results, but to compare the PSD of
$\delta L$ (visible in Figure \ref{LineDensity}) between the turbulent
and laminar simulations. To this end we use a `virtual' probe a
1mm$^3$ cube centred at the origin. After the initial transient we
compute the total vortex line density, within this region, to generate
a time series of $\delta L$. Using Welch's method a PSD is computed
from this signal. In Figure \ref{LPSD} (left) we compare the results
of the laminar and turbulent simulations at T=1.9~K (results at
T=1.6~K are consistent). A flat region, within the `inertial range' of
the turbulence is visible, with a power law decay at higher
frequencies. In Figure \ref{LPSD} (right) we show that changing the
location of the `probe' to the boundaries, or increasing the size of
the volume has no effect on the features of the PSD. We also show that
the scaling of the decay at high frequency is consistent with
$f^{-3}$.

The fact that the Reynolds number of the normal fluid is relatively
modest, as well as the fact that we take a frozen snapshot, could be
responsible for the fact that we observed no difference in the nature
of the PSDs between the turbulent and laminar simulations. In contrast
at high Reynolds numbers in mechanically pumped coflow \cite{Roche07}
a PSD scaling as $f^{-5/3}$ was observed. This result can be explained
as randomly oriented vortex lines are advected like passive scalars
which explains a $f^{-5/3}$ scaling in Kolmogorov turbulence
\cite{BaggaleyPRL} . Further work is required to determine whether a
Kolmogorov scaling is observed in counterflow turbulence when the
Reynolds number of the normal fluid is large.

\section{Conclusions}\label{sec:conc}

To summarise we have presented clear evidence that turbulence in the
normal fluid can sustain a substantially higher vortex line density
for a given mean flow rate in thermal counterflow.  We computed the scaling coefficient $\gamma$, 
and found a marked increase in its value when the normal fluid is turbulent. This offers the first numerical 
evidence that the transition from TI to TII turbulent
states is due to turbulence in the normal fluid
\cite{Tough1982,Melotte1998}.  Turbulence in the normal fluid leads to
higher curvature along the vortices, particularly in the centre of the
channel, this allows more energy to be extracted from the normal fluid
through the Donnelly-Glaberson instability, and supports the observed
increase in vortex line density.  
Through quantification of the polarization of counterflow generated vortex configurations we find that the tangle has a degree of negative polarization; the amount of polarisation appears to be correlated with the vortex line density. We offer evidence that vortex reconnections are a possible cause of negative polarization.
We also examined the power spectral
densities of the fluctuations of the vortex line density; our findings
are consistent with experimental and theoretical results. At low
Reynolds numbers considered in this study we observe no discernible
difference between the PSDs computed from the laminar and turbulent
simulations. In future studies we plan to investigate the effect of
increasing flow rate with a time-dependent turbulent counterflow.

\begin{acknowledgments}
AWB acknowledges the support of the Carnegie Trust, and fruitful
discussions with Carlo Barenghi and Sergey Nazarenko. We are grateful for the comments of the anonymous referees who's comments added to the scope and depth of the paper.
\end{acknowledgments}
\bibliographystyle{aip}
\bibliography{biblio} 
\end{document}